\documentclass[12pt,a4paper,english,preprint]{revtex4}
\usepackage[T1]{fontenc}
\usepackage[latin1]{inputenc}
\usepackage{array}
\usepackage{amsmath}
\usepackage{graphicx}
\usepackage{setspace}
\usepackage{amssymb}

\makeatletter

\providecommand{\tabularnewline}{\\}

\usepackage{babel}
\makeatother
\begin{document}

\title{Probing Little Higgs Model in $e^{+}e^{-}\rightarrow\nu\bar{\nu}\gamma$
Process}

\author{T. Aliev}

\email{taliev@metu.edu.tr}

\affiliation{Middle East Technical University, Faculty of Arts and Science, Department
of Physics, Ankara, Turkey.}

\author{O. \c{C}ak{\i}r}

\email{ocakir@science.ankara.edu.tr}

\affiliation{Ankara University, Faculty of Sciences, Department of Physics, 06100,
Tandogan, Ankara, Turkey.}

\begin{abstract}
We study the process $e^{+}e^{-}\rightarrow\nu\bar{\nu}\gamma$ to
search for its sensitivity to the extra gauge bosons $Z_{2}$, $Z_{3}$
and $W_{2}^{\pm}$ which are suggested by the little Higgs models.
We find that the ILC with $\sqrt{s}=0.5$ TeV and CLIC with $\sqrt{s}=3$
TeV cover different regions of the LHM parameters. We show that this
channel can provide accurate determination of the parameters, complementary
to measurements of the extra gauge bosons at the coming LHC experiments.
\end{abstract}
\maketitle

\section{Introduction}

Despite the impressive success of the Standard Model (SM) in describing
all existing experimental data at currently available energies, it
contains many unsolved problems. For example, origin of the fermion
mass, origin of the CP violation, hierarchy problems, etc. Therefore,
it is commonly believed that SM is low energy manifestaion of more
fundamental theory. In order to solve the hierarchy and fine-tuning
problems between the electroweak scale and the Planck scale, new physics
at the TeV scale is expected. In coming years the Large Hadron Collider
(LHC) and later International Linear Collider (ILC) will provide us
detailed information about the electroweak symmetry breaking and the
origin of the hierarchy of fermion masses and CP-violating interactions.
The supersymmetry introduces an extended space-time symmetry and removes
the quadratically divergent corrections due to the superpartners of
fermions and bosons. Extra dimensions reinterpret the problem completely
by lowering the fundamental Planck scale. Technicolor theories introduce
new strong dynamics at scale not much above the electroweak scale,
thus defer the hierarchy problem. Among the most popular non-supersymmetric
model for solving hierarchy problem in so-called little Higgs model
\cite{Arkani02} (see for example \cite{Han03} and references therein).
It is expected that the global symmetry breaking scale $\lesssim10$
TeV in order for the little Higgs model to be relevant for the hierarchy.
The little Higgs model solves the problem at one-loop level by eliminating
the quadratic divergencies via the presence of a partially broken
global symmetry $SU(5)$. The masses of these gauge bosons are expected
to be order of global symmetry breaking scale $f$ for $SU(5)\to SO(5)$.
In other words, the new heavy particles in this model cancel the quadratic
divergencies in question. The subgroup $[SU(2)\times U(1)]^{2}$ is
also broken into $SU(2)_{L}\times U(1)_{Y}$ group of the SM at the
scale $f$ of a few TeV and then $U(1)_{em}$ at the Fermi scale $v\simeq246$
GeV. The minimal type is the 'Littlest Higgs Model' (LHM), in addition
to the SM particles new charged heavy vector bosons $W_{2}^{\pm}$
(or heavy $W_{H}^{\pm}$), two neutral vector bosons $Z_{2}$ (or
heavy $Z_{H}$) and $Z_{3}$ (or heavy photon $A_{H}$), a heavy top
quark ($T$) and a triplet of scalar heavy particles ($\phi^{\pm},\phi^{0}$)
are present.

Since the LHM predicts many new particles, then search of these particles
usually are performed in two different way: i) via their indirect
effects, i.e. these particles new at loop and change SM predictions
on flavor changing neutral current processes (FCNC), ii) their direct
productions in high energy colliders. The relevant scale $f$ of new
physics must be $\gtrsim2-3$ TeV in order to be consistent with the
electroweak precision data \cite{Csaki03,Blanke07,Kai07,Conley05}.
Consequence of littlest Higgs model in rare FCNC $B$ and $K$ decays
comprehensively studied in the works \cite{Blanke07}. Direct productions
of new particles in high energy colliders are discussed in the works
\cite{Kai07}. The direct production of new heavy gauge bosons are
kinematically limited by the available center of mass energy of the
present colliders. At the Large Hadron Collider (LHC), the possible
signals of extra gauge bosons would show up through peaks in the invariant
mass distributions of their decay products \cite{Azuelos04}.

In present work, we study the indirect effects of extra gauge bosons
in the cross sections of the process $e^{+}e^{-}\rightarrow\nu\bar{\nu}\gamma$
at high energy linear $e^{+}e^{-}$ colliders; namely, International
Linear Collider (ILC) \cite{ILC} and Compact Linear Collider (CLIC)
\cite{CLIC}. In additon to the limits from hadron colliders, an improvement
on the sensitivity of the physical observables will be reached at
future $e^{+}e^{-}$linear colliders. Finally, we discuss how accurately
the LHM parameters will be measurable at the ILC and CLIC.

\section{Theoretical framework}

The process $e^{+}e^{-}\to\nu\bar{\nu\gamma}$ is widely discussed
in connection of determination of number of neutrino \cite{Ma78}
and understanding dynamics of stellar processes. Before discussion
of the $e^{+}e^{-}\to\nu\bar{\nu}\gamma$ process in the LHM few illuminating
remarks about main ingredients of the LHM are in order. In the little
Higgs model in addition to the standard $W^{\pm}$ and $Z$ boson
contributions there are contributions coming from new heavy vector
bosons, i.e. from extended gauge sector. The kinetic term of the scalar
field $\Sigma$ in lagrangian has the form \cite{Arkani02}

\begin{equation}
L=\frac{f^{2}}{8}Tr\left|D_{\mu}\Sigma\right|^{2}\label{eq:1}\end{equation}
with the covariant derivative of the scalar $\Sigma$ field

\begin{equation}
D_{\mu}\Sigma=\partial_{\mu}\Sigma-i{\displaystyle \sum_{k=1}^{2}\left[g_{k}\left(W_{k}\Sigma+\Sigma W_{k}^{T}\right)+g_{k}^{'}\left(B_{k}\Sigma+\Sigma B_{k}^{T}\right)\right]}\label{eq:2}\end{equation}
where $g_{k}$ and $g'_{k}$ are the coupling constants related to
the gauge fields $W_{k}$ and $B_{k}$. The mixing angles $s$ and
$s'$, $s=g_{2}/\sqrt{g_{1}^{2}+g_{2}^{2}}$ and $s^{'}=g_{2}^{'}/\sqrt{g_{1}^{'2}+g_{2}^{'2}}$,
relates the coupling strengths of the two $SU(2)\times U(1)$ gauge
groups. Relations between gauge bosons in weak and mass eigenstates
similar to the SM case; namely

\begin{equation}
\left(\begin{array}{c}
W\\
W'\end{array}\right)=\left(\begin{array}{cc}
s & c\\
-c & s\end{array}\right)\left(\begin{array}{c}
W_{1}\\
W_{2}\end{array}\right)\;,\;\left(\begin{array}{c}
B\\
B'\end{array}\right)=\left(\begin{array}{cc}
s' & c'\\
-c' & s'\end{array}\right)\left(\begin{array}{c}
B_{1}\\
B_{2}\end{array}\right)\label{eq:3}\end{equation}
where the $W$ and $B$ are the gauge boson states associated with
the generators of $SU(2)$ and $U(1)$ of the SM. The $W'$ and $B'$
are the massive gauge bosons with their masses $m_{W'}=gf/2sc$ and
$m_{B'}=g'f/2\sqrt{5}s'c'$. Here $s,s'(c,c')$ represent the sine
(cosine) of two mixing angles. After electroweak symmetry breaking
all the light and heavy gauge bosons are obtained, and they include
$Z_{1},W_{1}^{\pm},\gamma$ of the SM and $W_{2}^{\pm},Z_{2},Z_{3}$
of the LHM.

\begin{figure}
\begin{center}\includegraphics[%
  scale=0.8]{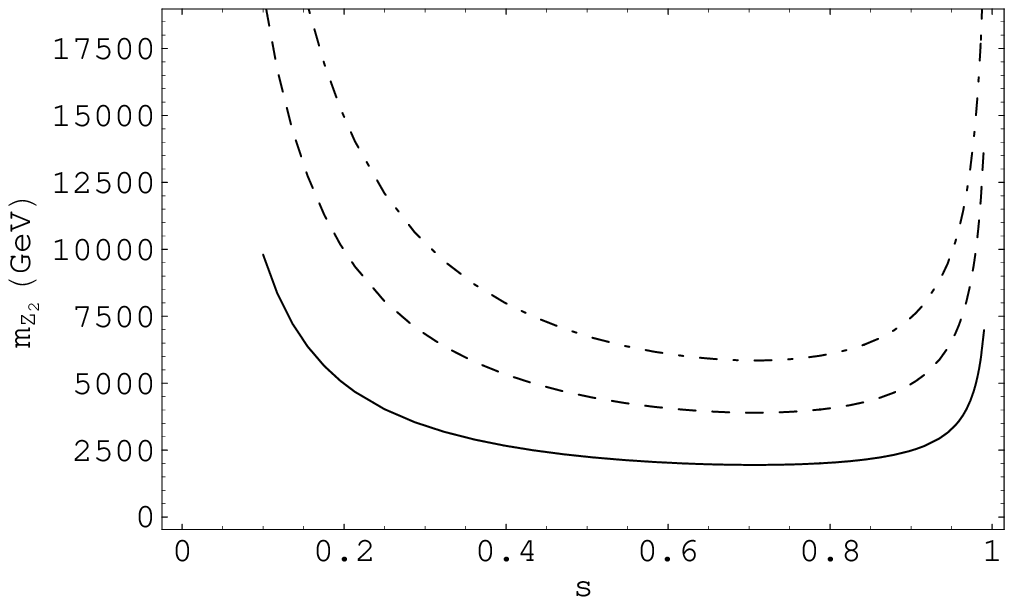} \includegraphics[%
  scale=0.8]{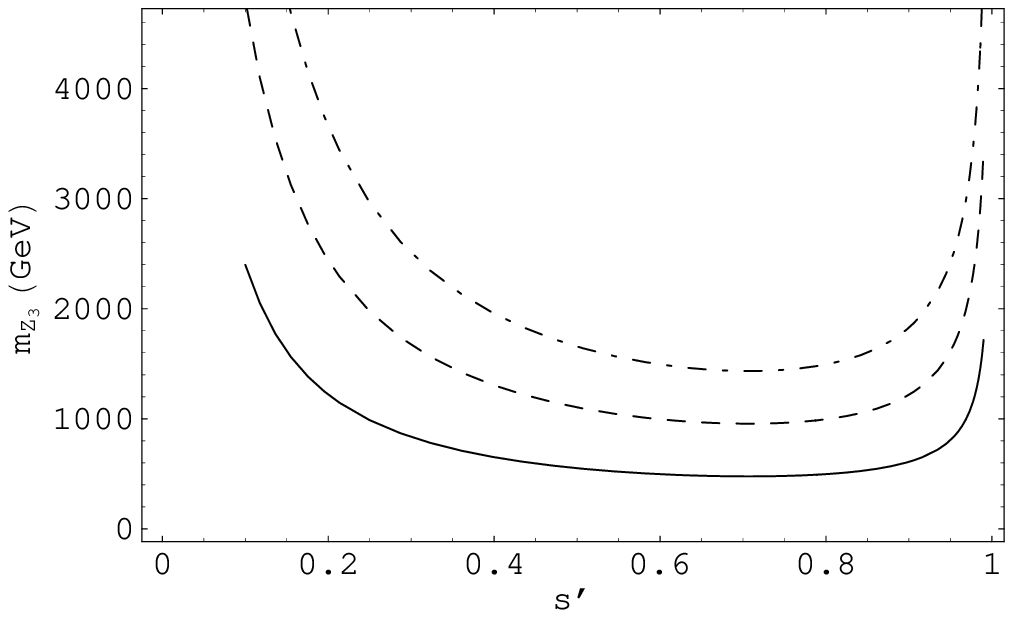}\end{center}

\caption{Heavy gauge boson masses $m_{Z_{2}}$ (left) and $m_{Z_{3}}$ (right),
depending on the mixing $s$ (where $s'=0.5$) and $s'$ (where $s=0.5$)
for different scale $f=3$ TeV (solid line), $f=6$ TeV (dashed line)
and $f=9$ TeV (dot-dashed line).\label{fig1}}
\end{figure}

The masses of the new heavy gauge bosons in the LHM to the order of
$\mathcal{O}(v^{2}/f^{2})$ are given by following expressions \cite{Han03}:

\begin{eqnarray}
m_{Z_{1}} & = & m_{Z}\left[1-\frac{v^{2}}{f^{2}}\left(\frac{1}{6}+\frac{1}{4}(c^{2}-s^{2})^{2}+\frac{5}{4}(c'^{2}-s'^{2})^{2}+8\frac{v'^{2}}{v^{2}}\right)\right]^{1/2}\label{eq:4}\\
m_{Z_{2}} & = & m_{W}\left(\frac{f^{2}}{s^{2}c^{2}v^{2}}-1-\frac{x_{H}s_{W}^{2}}{s'^{2}c'^{2}c_{W}^{2}}\right)^{1/2}\label{eq:5}\\
m_{Z_{3}} & = & m_{Z}s_{W}\left(\frac{f^{2}}{5s'^{2}c'^{2}v^{2}}-1+\frac{x_{H}c_{W}^{2}}{4s^{2}c^{2}s_{W}^{2}}\right)^{1/2}\label{eq:6}\\
m_{W_{1}} & = & m_{W}\left[1-\frac{v^{2}}{f^{2}}\left(\frac{1}{6}+\frac{1}{4}(c^{2}-s^{2})^{2}\right)+4\frac{v'^{2}}{v^{2}}\right]^{1/2}\label{eq:7}\\
m_{W_{2}} & = & m_{W}\left(\frac{f^{2}}{s^{2}c^{2}v^{2}}-1\right)^{1/2}\label{eq:8}\end{eqnarray}
where $m_{Z}$ and $m_{W}$ are the SM gauge boson masses and $c_{W}(s_{W})$
denotes the cosine (sine) of the weak mixing angle. Here $x_{H}$
characterizes the mixing between $B'$ and $W'$ in the $Z_{2}$ and
$Z_{3}$ eigenstates and depends on gauge couplings. As can be seen
from Fig. \ref{fig1}, the masses of new neutral gauge bosons $Z_{2}(Z_{3})$
strongly depends on $s(s')$. From equations (5) and (6) we obtain
the ratio satisfying $m_{Z_{3}}/m_{Z_{2}}\simeq0.25$ for some ranges
of the parameters $s,s'$. Fig. \ref{fig1} reflects this property
and the mass of the $Z_{3}$ boson remains below 1 TeV for a wide
range of the parameter $s'$. We may note that $Z_{3}$ is much lighter
than $Z_{2}$ and could be searched at ILC energies. If ILC does not
discover the boson $Z_{3}$ it is possible to put a lower bound on
the scale $f\gtrsim3$ TeV.

\begin{table}

\caption{Neutral and charged gauge boson-fermion couplings in the little Higgs
model. Last line denote $W_{1(2)}^{+}W_{1(2)}^{-}\gamma$ couplings.
\label{table1}}

\tiny\begin{tabular}{|c|>{\centering}p{10cm}|c|}
\hline
Particles&
\begin{singlespace}
$g_{V}$\end{singlespace}
&
$g_{A}$\tabularnewline
\hline
$Z_{1}\nu\bar{\nu}$&
$\frac{g}{2c_{W}}\left\{ \frac{1}{2}-\frac{v^{2}}{f^{2}}\left[c_{W}x_{Z}^{W'}\frac{c}{2s}+\frac{s_{W}x_{Z}^{B'}}{s'c'}\left(y_{e}-\frac{4}{5}+\frac{c'^{2}}{2}\right)\right]\right\} $&
$-g_{V}$\tabularnewline
\hline
$Z_{2}\nu\bar{\nu}$&
$gc/4s$&
$-g_{V}$\tabularnewline
\hline
$Z_{3}\nu\bar{\nu}$&
$\frac{g'}{2s'c'}\left(y_{e}-\frac{4}{5}+\frac{c'^{2}}{2}\right)$&
$-g_{V}$\tabularnewline
\hline
$Z_{1}e^{-}e^{+}$&
$\frac{g}{2c_{W}}\left\{ -\frac{1}{2}+2s_{W}^{2}-\frac{v^{2}}{f^{2}}\left[-c_{W}x_{Z}^{W'}\frac{c}{2s}+\frac{s_{W}x_{Z}^{B'}}{s'c'}\left(2y_{e}-\frac{9}{5}+\frac{3c'^{2}}{2}\right)\right]\right\} $&
$\frac{g}{2c_{W}}\left\{ \frac{1}{2}-\frac{v^{2}}{f^{2}}\left[c_{W}x_{Z}^{W'}\frac{c}{2s}+\frac{s_{W}x_{Z}^{B'}}{s'c'}\left(-\frac{1}{5}+\frac{c'^{2}}{2}\right)\right]\right\} $\tabularnewline
\hline
$Z_{2}e^{-}e^{+}$&
$-gc/4s$&
$-g_{V}$\tabularnewline
\hline
$Z_{3}e^{-}e^{+}$&
$\frac{g'}{2s'c'}\left(2y_{e}-\frac{9}{5}+\frac{3c'^{2}}{2}\right)$&
$\frac{g'}{2s'c'}\left(-\frac{1}{5}+\frac{c'^{2}}{2}\right)$\tabularnewline
\hline
&
\multicolumn{2}{c|}{Coupling $g_{W}$}\tabularnewline
\hline
$W_{1}^{+}e^{-}\bar{\nu}$&
\multicolumn{2}{c|}{$\frac{g}{2\sqrt{2}}\left[1-\frac{v^{2}}{2f^{2}}c^{2}(c^{2}-s^{2})\right]$}\tabularnewline
\hline
$W_{2}^{+}e^{-}\bar{\nu}$&
\multicolumn{2}{c|}{$-\frac{g}{2\sqrt{2}}\frac{c}{s}\left[1+\frac{v^{2}}{2f^{2}}s^{2}(c^{2}-s^{2})\right]$}\tabularnewline
\hline
\end{tabular}
\end{table}

\begin{figure}
\begin{center}\includegraphics[%
  scale=0.8]{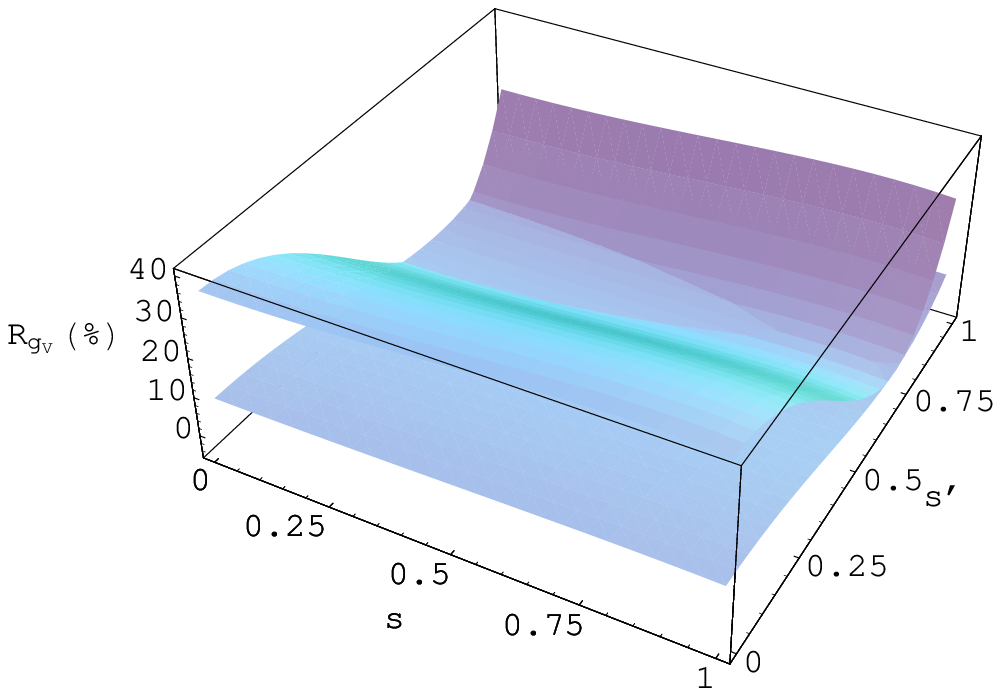}\includegraphics[%
  scale=0.8]{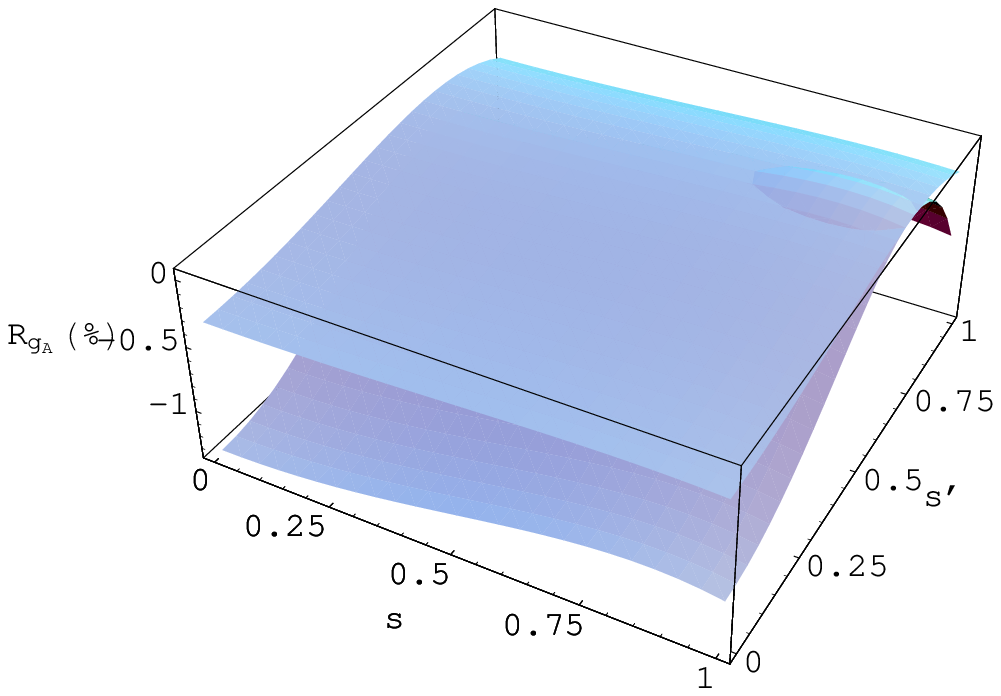}\end{center}

\caption{The relative changes $R_{g_{V}}(\%)$ and $R_{g_{A}}(\%)$ of $Z_{1}e^{+}e^{-}$
vector $g_{V}$ and axial-vector $g_{A}$ couplings from the SM values
depending on $s$ and $s'$ taking the scale $f=3$ TeV (upper on
left panel, lower on right panel) and $f=6$ TeV (lower on left panel,
upper on right panel). \label{fig2}}
\end{figure}

The coupling between gauge bosons and fermions can be written in the
form $-i\gamma^{\mu}(g_{V}+g_{A}\gamma^{5})$. The couplings $g_{V}$
and $g_{A}$ also depend on the mixing parameter $s,s'$ and the scale
$f$. The expressions for these couplings are given in Table \ref{table1}.
In order to see how $Z_{1}e^{+}e^{-}$ vector and axial-vector couplings
change from their SM values we give a 3D plot as shown in Fig. \ref{fig2}.
We find that the relative changes in $g_{V}$ is much greater than
that for $g_{A}$ for the values of $s'$ near the endpoints. It is
possible to set a bound on $s$ and $s'$ by demanding these couplings
remain perturbative, and hence one obtain a limit $s,s'>0.1$. As
can be seen from Table \ref{table1}, $Z_{3}l\bar{l}$ coupling vanishes
for $c'=\sqrt{2/5}$ once given $y_{e}=0.6$.

The couplings of the $Z_{1}$ boson and $W_{1}$ boson to the SM leptons
are subject to corrections in the LHM. Using their couplings shown
in Table \ref{table1} one obtains for the $Z_{1}$ total decay width
and $W_{1}$ boson mass up to corrections proportional to $\mathcal{O}(v^{2}/f^{2})$:
$\Gamma_{Z_{1}}\simeq\Gamma_{Z}(1+1.7v^{2}/f^{2})$ and $m_{W_{1}}\simeq m_{W}(1+0.89v^{2}/f^{2})$,
leading to the comment that $f>8$ TeV even for small $c'$. Since
there is some partial cancellations, in fact as a general guide we
take $v/f\lesssim0.1$. We present the decay widths of $Z_{2}$ and
$W_{2}^{\pm}$ bosons which we need in the calculation of the cross
section for process $e^{+}e^{-}\to\nu\bar{\nu}\gamma$. The decay
of heavy gauge boson $Z_{2}$ include leptonic, hadronic and gauge
boson channels to give the partial widths of the form \cite{Han03}

\[
\Gamma(Z_{2}\rightarrow l^{+}l^{-})\simeq\frac{g^{2}\cot^{2}\theta}{96\pi}m_{Z_{2}},\qquad\Gamma(Z_{2}\rightarrow\bar{q}q)\simeq\frac{g^{2}\cot^{2}\theta}{32\pi}m_{Z_{2}}\]

\begin{equation}
\Gamma(Z_{2}\rightarrow Z_{1}h)\simeq\frac{g^{2}\cot^{2}2\theta}{192\pi}m_{Z_{2}},\qquad\Gamma(Z_{2}\rightarrow W_{1}^{+}W_{1}^{-})\simeq\frac{g^{2}\cot^{2}2\theta}{192\pi}m_{Z_{2}}\label{eq:9}\end{equation}
where we neglect the {\small corrections} from the $v/f$ terms and
the final state masses. The partial decay widths for the $W_{2}^{\pm}$
bosons can be obtained from (\ref{eq:9}) using the isospin symmetry,
as follows

\[
\Gamma(W_{2}^{\pm}\rightarrow l^{\pm}\nu)\simeq\frac{g^{2}\cot^{2}\theta}{48\pi}m_{W_{2}},\qquad\Gamma(W_{2}^{\pm}\rightarrow\bar{q}'q)\simeq\frac{g^{2}\cot^{2}\theta}{16\pi}m_{W_{2}}\]

\begin{equation}
\Gamma(W_{2}^{\pm}\rightarrow W_{1}^{\pm}h)\simeq\frac{g^{2}\cot^{2}2\theta}{192\pi}m_{Z_{2}},\qquad\Gamma(W_{2}^{\pm}\rightarrow W_{1}^{\pm}Z_{1})\simeq\frac{g^{2}\cot^{2}2\theta}{192\pi}m_{Z_{2}}\label{eq:10}\end{equation}

The gauge boson $Z_{3}$ is assumed to be light and could be explored
at future colliders. Similarly, its decay width can be obtained from
(1) by replacing $g\to g'$ and $\theta\to\theta'$.

\begin{figure}
\begin{center}\includegraphics{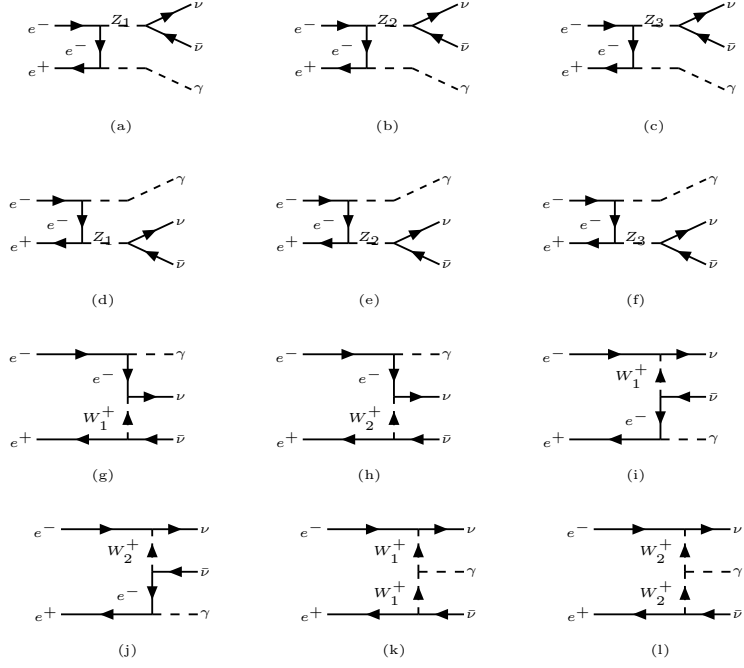}\end{center}

\caption{The Feynman diagrams contributing to the process $e^{+}e^{-}\rightarrow\nu\bar{\nu}\gamma$.\label{fig3}}
\end{figure}

After these preliminary remarks, let we consider the process $e^{-}(p_{1})e^{+}(p_{2})\rightarrow\nu(k_{1})\bar{\nu}(k_{2})\gamma(k)$
in LHM for which relevant diagrams are presented in Fig. \ref{fig3}.
In the SM, this process proceeds via s-channel $Z$ and t-channel
$W^{\pm}$ exchange with the photon being radiated from the initial
charged paticles. In the LHM models this process has also contributions
from both s-channel $Z_{2},Z_{3}$ and t-channel $W_{2}^{\pm}$ exchange.
We implement all relevant vertices in the CalcHEP \cite{Pukhov99}
in the framework of the littlest Higgs model. The amplitudes for the
diagrams Fig. \ref{fig3}(a-c) are given by

\begin{eqnarray}
M_{1} & = & \sum_{a=1}^{3}\bar{u}(k_{1})[-i\gamma^{\mu}(g_{V}^{a(\nu)}+g_{A}^{a(\nu)}\gamma_{5})]v(k_{2})\left[\frac{-i(g_{\mu\nu}-q_{1\mu}q_{1\nu}/m_{Z_{a}}^{2})}{q_{1}^{2}-m_{Z_{a}}^{2}+im_{Z_{a}}\Gamma_{Z_{a}}}\right]\nonumber \\
 &  & \bar{v}(p_{2})(ig_{e}\not\varepsilon)\left[\frac{i(\not q+m_{e})}{q^{2}-m_{e}^{2}}\right](-i\gamma^{\nu})(g_{V}^{a(e)}+g_{A}^{a(e)}\gamma_{5})u(p_{1})\label{eq:11}\end{eqnarray}

where $q_{1}=k_{1}+k_{2}$, $q=k-p_{2}$ and $\varepsilon_{\mu}$
is the photon polarisation four-vector. The amplitudes for Fig. \ref{fig3}(d-f)
are given by

\begin{eqnarray}
M_{2} & = & \sum_{a=1}^{3}\bar{u}(k_{1})[-i\gamma^{\mu}(g_{V}^{a(\nu)}+g_{A}^{a(\nu)}\gamma_{5})]v(k_{2})\left[\frac{-i(g_{\mu\nu}-q_{1\mu}q_{1\nu}/m_{Z_{a}}^{2})}{q_{1}^{2}-m_{Z_{a}}^{2}+im_{Z_{a}}\Gamma_{Z_{a}}}\right]\nonumber \\
 &  & \bar{v}(p_{2})(-i\gamma^{\nu})(g_{V}^{a(e)}+g_{A}^{a(e)}\gamma_{5})\left[\frac{i(\not q'+m)}{q'^{2}-m_{e}^{2}}\right](ig_{e}\not\varepsilon)u(p_{1})\label{eq:12}\end{eqnarray}

where $q'=p_{1}-k$. The amplitudes for Fig. \ref{fig3}(g,h) are
given by

\begin{eqnarray}
M_{3} & = & \sum_{b=1}^{2}\bar{u}(k_{1})(-ig_{V}^{b}\gamma^{\mu})(1-\gamma_{5})\frac{i(\not q+m_{e})}{q'^{2}-m_{e}^{2}}](ig_{e}\not\varepsilon)u(p_{1})\nonumber \\
 &  & \left[\frac{-i(g_{\mu\nu}-q_{3\mu}q_{3\nu}/m_{W_{b}}^{2})}{q_{3}^{2}-m_{W_{b}}^{2}+im_{W_{b}}\Gamma_{W_{b}}}\right]\bar{v}(p_{2})(-ig_{V}^{b}\gamma^{\nu})(1-\gamma_{5})v(k_{2})\label{eq:13}\end{eqnarray}

where $q_{3}=p_{2}-k_{2}$. The amplitudes for Fig. \ref{fig3}(i,j)
are given by

\begin{eqnarray}
M_{4} & = & \sum_{b=1}^{2}\bar{u}(k_{1})(-ig_{V}^{b}\gamma^{\mu})(1-\gamma_{5})u(p_{1})\left[\frac{-i(g_{\mu\nu}-q_{4\mu}q_{4\nu}/m_{W_{b}}^{2})}{q_{4}^{2}-m_{W_{b}}^{2}+im_{W_{b}}\Gamma_{W_{b}}}\right]\nonumber \\
 &  & \bar{v}(p_{2})(ig_{e}\not\varepsilon)\left[\frac{i(\not q+m)}{q^{2}-m_{e}^{2}}\right](-ig_{V}^{b}\gamma^{\nu})(1-\gamma_{5})v(k_{2})\label{eq:14}\end{eqnarray}

where $q_{4}=k_{1}-p_{1}$. The amplitudes for Fig. \ref{fig3}(k,l)
are given by

\begin{eqnarray}
M_{5} & = & \sum_{b=1}^{2}\bar{u}(k_{1})(-ig_{V}^{b}\gamma_{\mu})(1-\gamma_{5})u(p_{1})\left[\frac{-i(g^{\mu\mu'}-q_{4}^{\mu}q_{4}^{\mu'}/m_{W_{b}}^{2})}{q_{4}^{2}-m_{W_{b}}^{2}+im_{W_{b}}\Gamma_{W_{b}}}\right]\nonumber \\
 &  & ig_{e}(g_{\nu'\lambda}(q_{3}+k)_{\mu'}+g_{\lambda\mu'}(-k+q_{4})_{\nu'}+g_{\mu'\nu'}(-q_{4}-q_{3})_{\lambda})\varepsilon^{\lambda}\nonumber \\
 &  & \left[\frac{-i(g^{\nu\nu'}-q_{3}^{\nu}q_{3}^{\nu'}/m_{W_{b}}^{2})}{q_{3}^{2}-m_{W_{b}}^{2}+im_{W_{b}}\Gamma_{W_{b}}}\right]\bar{v}(p_{2})(-ig_{V}^{b}\gamma_{\nu})(1-\gamma_{5})v(k_{2})\label{eq:15}\end{eqnarray}

\section{Numerical Results}

\begin{figure}
\begin{center}\includegraphics{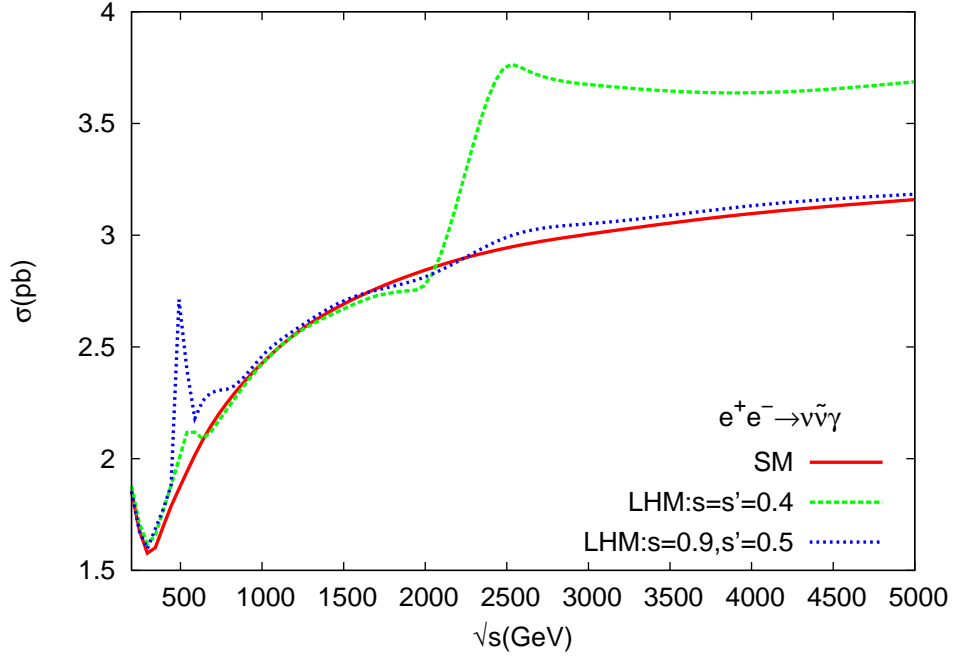}\end{center}

\caption{The total cross section in pb versus center of mass energy $\sqrt{s}$.
For the LHM model we take two different points for $s,s'$ and $v/f=0.1$.
\label{fig4}}
\end{figure}

We will interest the differential cross sections over the kinematic
observables of the photon energy $E_{\gamma}$ and its angle relative
to incident electron direction, respectively. The double differential
cross section of the considered process is given by

\begin{equation}
\frac{d\sigma}{dE_{\gamma}d\cos\theta_{\gamma}}=\frac{\left|M\right|^{2}E_{\gamma}}{128\pi^{3}s}\label{eq:16}\end{equation}
where the amplitude $M$ is the sum of above five amplitudes, $M_{1-5}$.
In order to remove the collinear singularities, when the photon is
emitted in the initial beam direction, we apply the initial kinematic
cuts: $E_{\gamma}>10$ GeV and $10^{o}<\theta_{e\gamma}<170^{o}$.
We may also impose a cut, $p_{T\gamma}>10$ GeV, on the transverse
momentum of photon to remove the large background from radiative Bhabba
scattering. Figure \ref{fig4} shows the total cross section for $e^{+}e^{-}\to\nu\bar{\nu}\gamma$
as a function of the center of mass energy $\sqrt{s}$ for the SM
and two different values of the LHM parameters $s$ and $s'$. Starting
from a center of mass energy just greater than the $Z$ mass, a minimum
around $\sqrt{s}\simeq300$ GeV occurs due to the SM $Z$-boson resonance
tail on the high energies. For different values of the parameters
$s,s'$ and $f$ the shape of the LHM curves changes leading to the
appearence/disappearence of the resonance peaks. For the proposed
energies and luminosities of the ILC and CLIC $e^{+}e^{-}$ colliders
we can well measure different extra gauge boson couplings for the
interested region of the parameters. In other words, preferably we
may search for $Z_{3}$ at ILC ($0.5-1$ TeV) energies and $Z_{2}$
at CLIC energies ($1-5$ TeV).

\begin{figure}
\begin{center}\includegraphics[%
  scale=0.6]{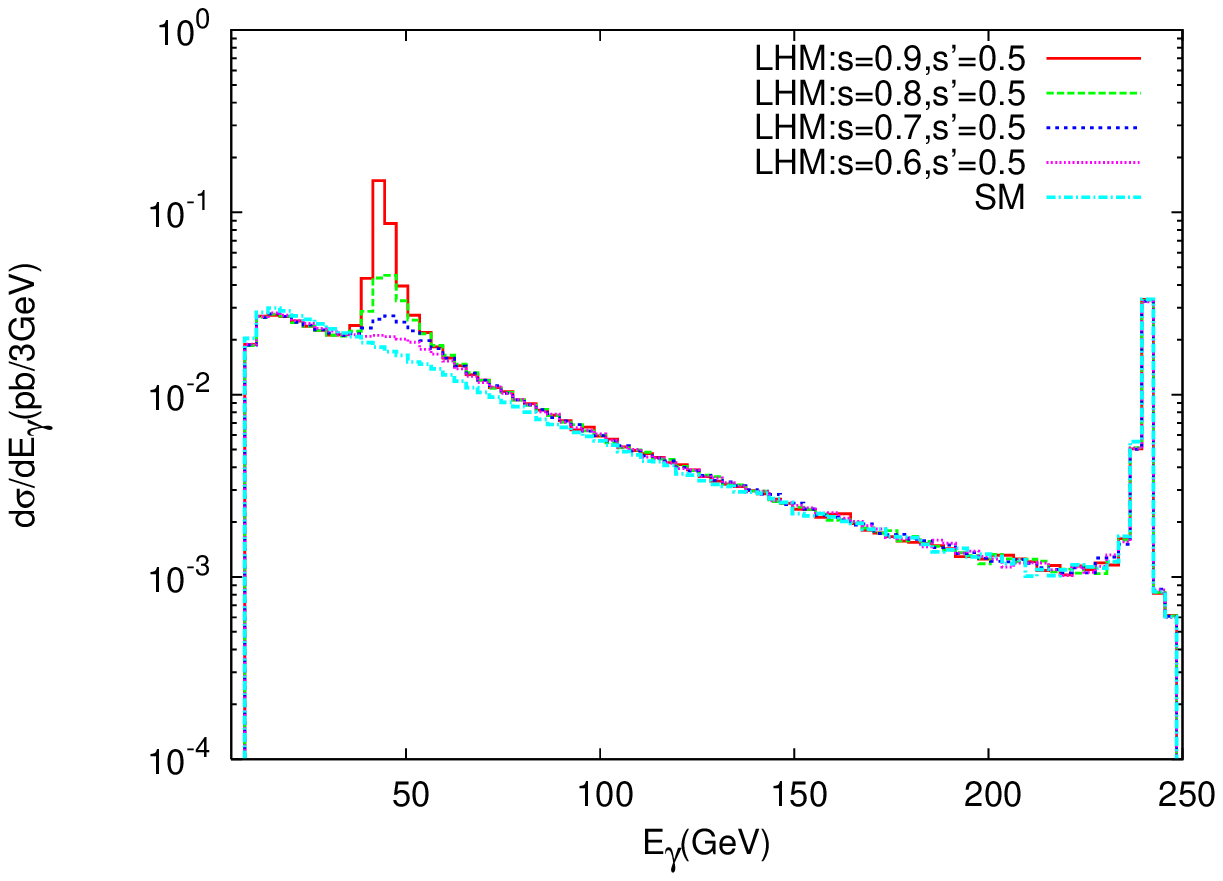} \includegraphics[%
  scale=0.6]{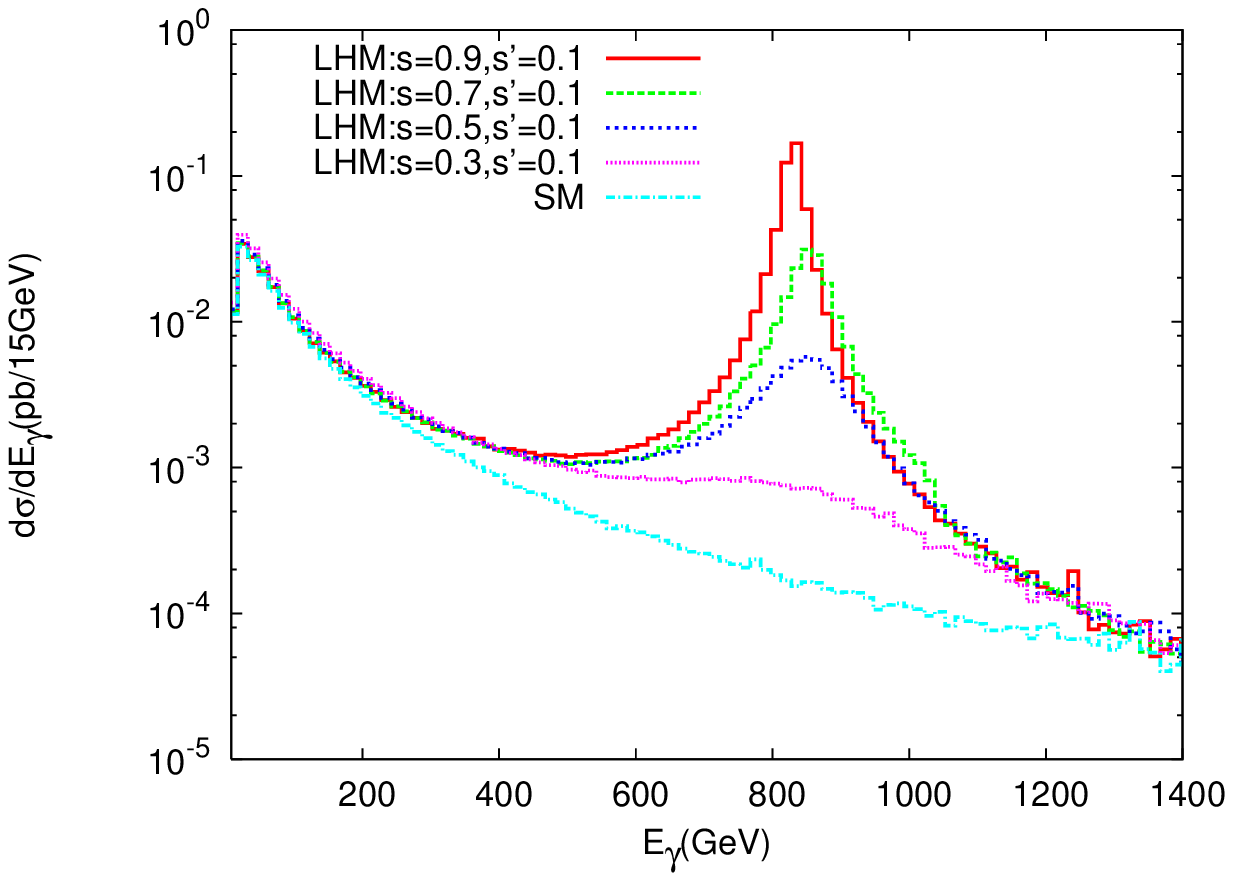}\end{center}

\caption{Diferential cross section versus photon energy at $\sqrt{s}=500$
GeV (left) and $\sqrt{s}=3000$ GeV (right) for $v/f=0.1$ and different
values of $s,s'$. \label{fig5}}
\end{figure}

\begin{table}

\caption{Masses and decay widths of neutral $(Z_{2,3})$ and charged $(W_{2}^{\pm})$
gauge bosons. Here we use $v/f=0.1$ and $y_{e}=0.6$.\label{table2}}

\begin{tabular}{|c|c|c|c|c|c|c|}
\hline
$\sin\theta/\sin\theta'$&
$m_{Z_{2}}$(GeV)&
$m_{Z_{3}}$(GeV)&
$m_{W_{2}}$(GeV)&
$\Gamma_{Z_{2}}$(GeV)&
$\Gamma_{Z_{3}}$(GeV)&
$\Gamma_{W_{2}}$(GeV)\tabularnewline
\hline
0.1/0.1&
8034.4&
1971.2&
8034.4&
27153.0&
6614.7&
26899.80\tabularnewline
\hline
0.3/0.1&
2787.0&
1971.7&
2792.4&
960.32&
693.95&
953.17\tabularnewline
\hline
0.4/0.1&
2138.4&
1972.7&
2179.6&
382.35&
370.61&
385.61\tabularnewline
\hline
0.5/0.3&
1843.9&
684.5&
1844.8&
187.99&
70.06&
186.09\tabularnewline
\hline
0.5/0.5&
1844.5&
451.7&
1844.8&
188.05&
45.90&
186.09\tabularnewline
\hline
0.5/0.9&
1844.1&
499.6&
1844.8&
188.01&
50.84&
186.09\tabularnewline
\hline
0.9/0.5&
2036.2&
452.6&
2036.6&
17.95&
3.65&
17.78\tabularnewline
\hline
0.9/0.9&
2036.3&
498.7&
2036.6&
17.95&
4.09&
17.78\tabularnewline
\hline
\end{tabular}
\end{table}

\begin{figure}
\includegraphics[%
  scale=0.6]{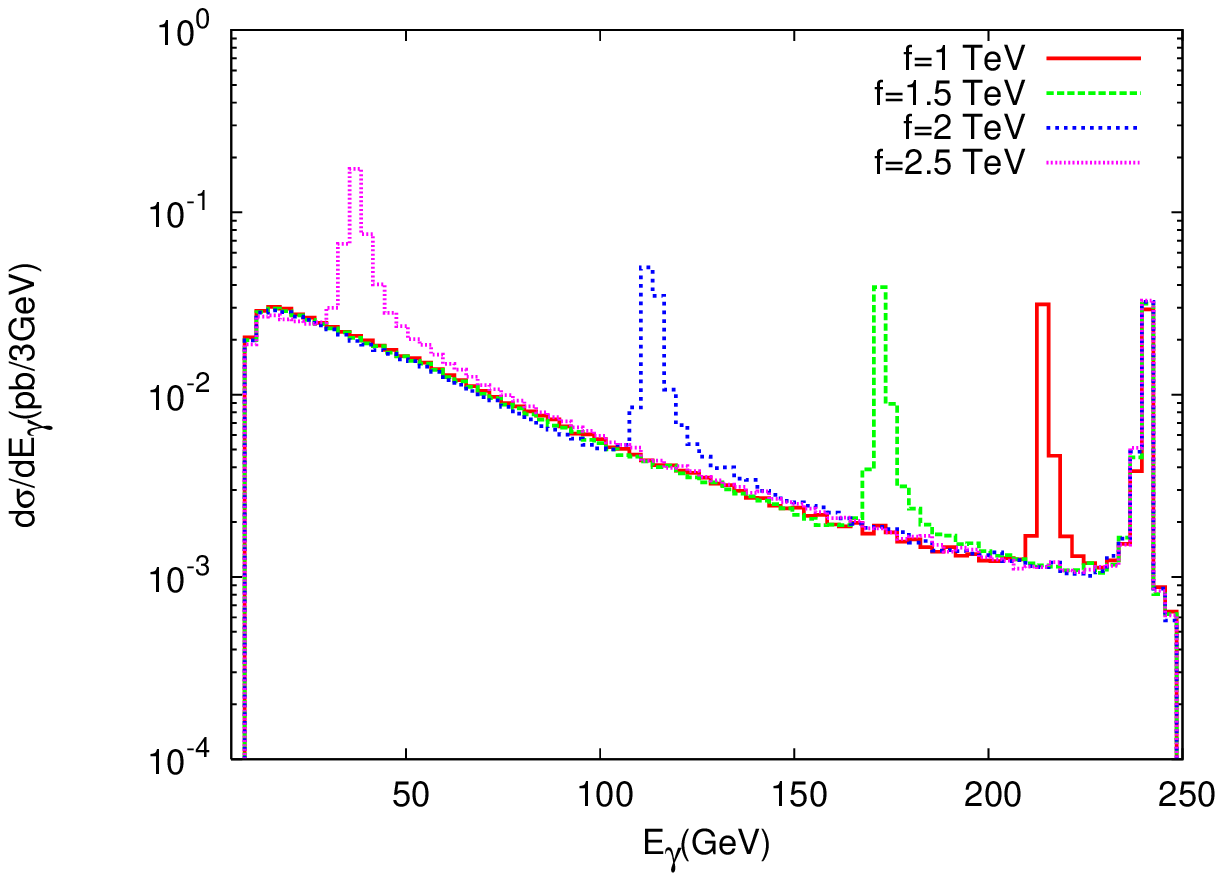} \includegraphics[%
  scale=0.6]{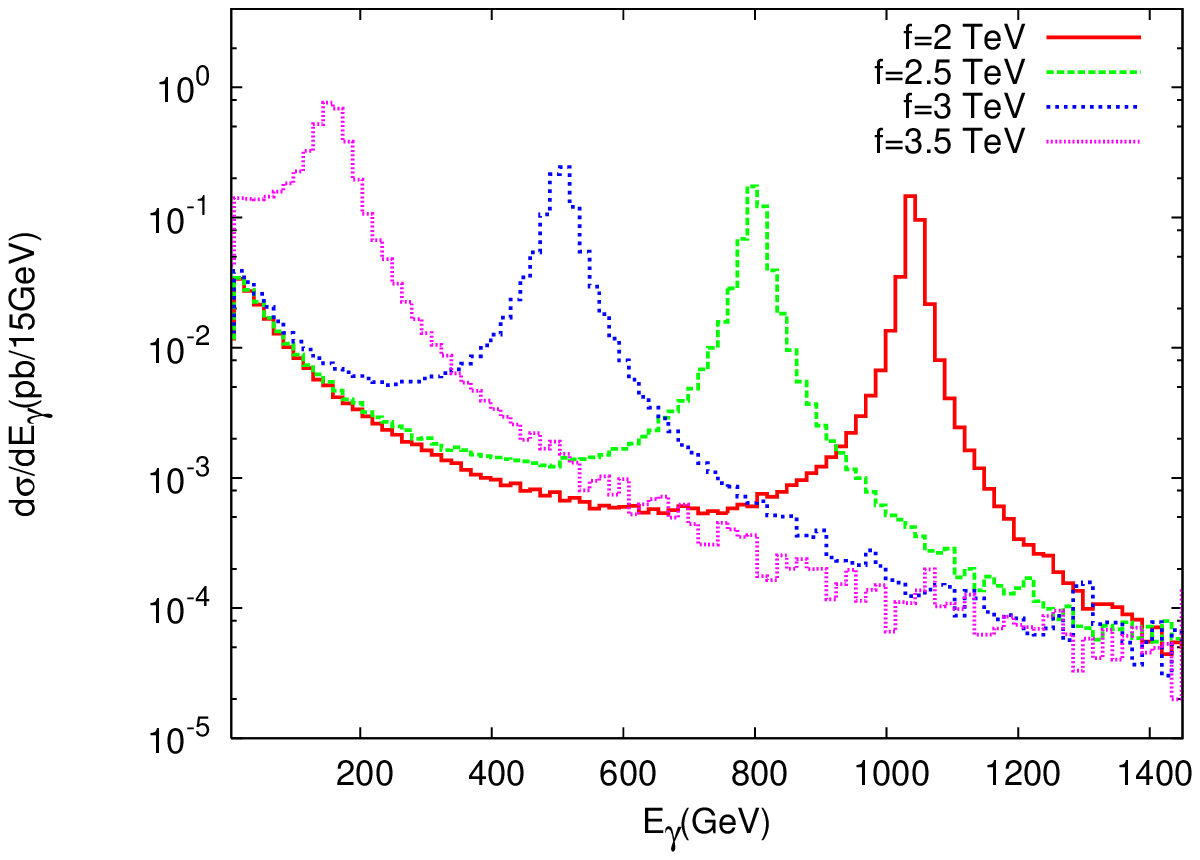}

\caption{Energy distribution of photon for different values of the scale $f$
at $\sqrt{s}=500$GeV (left) and $\sqrt{s}=3000$ GeV (right). The
sine of mixing angle is taken as $s=0.9$,$s'=0.5$ for the left plot,
and $s=0.9,s=0.1$ for the right plot.\label{fig6}}
\end{figure}

In table \ref{table3} and \ref{table4} we present the total cross
section for the process $e^{+}e^{-}\to\nu\bar{\nu}\gamma$ with both
signal and SM background. We find the total cross section (signal+background)
changes at most $\%44$ at $\sqrt{s}=0.5$ TeV for the interested
region of the parameters $s,s'$ with the scale $f=2.46$ TeV. There
is also a large contribution from extra gauge bosons, mainly $Z_{2}$,
for relatively small parameter $s'=0.1$ with a larger values of $s=0.9$
and the scale $f=3.5$ TeV at the center of mass energy $\sqrt{s}=3$
TeV as shown in table \ref{table4}. In order to see sensitivity of
the photon energy to new physics, in Fig. \ref{fig5} we plot the
differential cross section versus $E_{\gamma}$ by taking $v/f=0.1$
at the center of mass energy $\sqrt{s}=0.5$ TeV and $\sqrt{s}=3$
TeV, respectively. We see that for the value of parameter $s'=0.5$
the $Z_{3}$ resonance occurs as its magnitude strongly depends on
the values of $s$. The peak in the cross section due to $Z_{3}$
($Z_{2}$) boson shifts to the right as $s$ decrease. We see from
Figure \ref{fig5} that main contributions to the total cross section
(signal+background) comes from three regions, low energy region, resonance
region and the region due to radiative return to the $Z$ pole, where
$E_{\gamma}=\sqrt{s}(1-m_{Z}^{2}/s)/2\approx240$ GeV. The pole region
($\sim\sqrt{s}/2$) is quite insensitive to the new physics. The resonance
region for $Z_{3}$ occurs at $s'=0.5$ and $f\simeq1-3$ TeV. The
peak of the resonance shifts to lower photon energies (left) when
the scale $f$ increased as shown in Fig. \ref{fig6}. This is due
to the fact that as $f$ increases the extra gauge boson masses ($\propto f$)
also increase, as the resonance occurs there remains lower energy
delivered to the photon, i.e. the lower $E_{\gamma}$, the higher
the mass probed in the $Z_{i}$ propagator via $E_{\gamma}=\sqrt{s}(1-m_{Z_{i}}^{2}/s)/2$.
For a visible signal peak one can scan the parameter $f$ between
$\simeq1-3$ TeV at a collider energy of $\sqrt{s}=0.5$ TeV. At higher
center of mass energies such as $\sqrt{s}=3$ TeV this resonance scan
can be extended to upper values of the scale $f$ around $\simeq2-4$
TeV.

\begin{table}

\caption{The cross sections (in pb) for $e^{+}e^{-}\to\nu\bar{\nu}\gamma$
with $v/f=0.1$ at $\sqrt{s}=500$ GeV. The corresponding SM background
gives $\sigma_{B}=1.879$ pb. Here we applied the minimal cuts $E_{\gamma}>10$
GeV, $10^{o}<\theta_{13}<170^{o}$ and $p_{T}>10$ GeV.\label{table3}}

\begin{tabular}{|c|c|c|c|c|c|}
\hline
$\sin\theta$\textbackslash{}$\sin\theta'$&
0.1&
0.3&
0.5&
0.7&
0.9\tabularnewline
\hline
0.1&
1.9379&
1.9347&
1.9382&
1.9396&
1.9384\tabularnewline
\hline
0.3&
1.9662&
1.9701&
1.9035&
1.8919&
1.9041\tabularnewline
\hline
0.5&
1.9761&
2.0012&
1.9294&
1.8806&
1.9305\tabularnewline
\hline
0.7&
1.9755&
1.9983&
2.0394&
1.8905&
1.9583\tabularnewline
\hline
0.9&
1.9606&
1.9915&
2.7090&
1.8878&
1.9668\tabularnewline
\hline
\end{tabular}
\end{table}

\begin{figure}
\includegraphics[%
  scale=0.6]{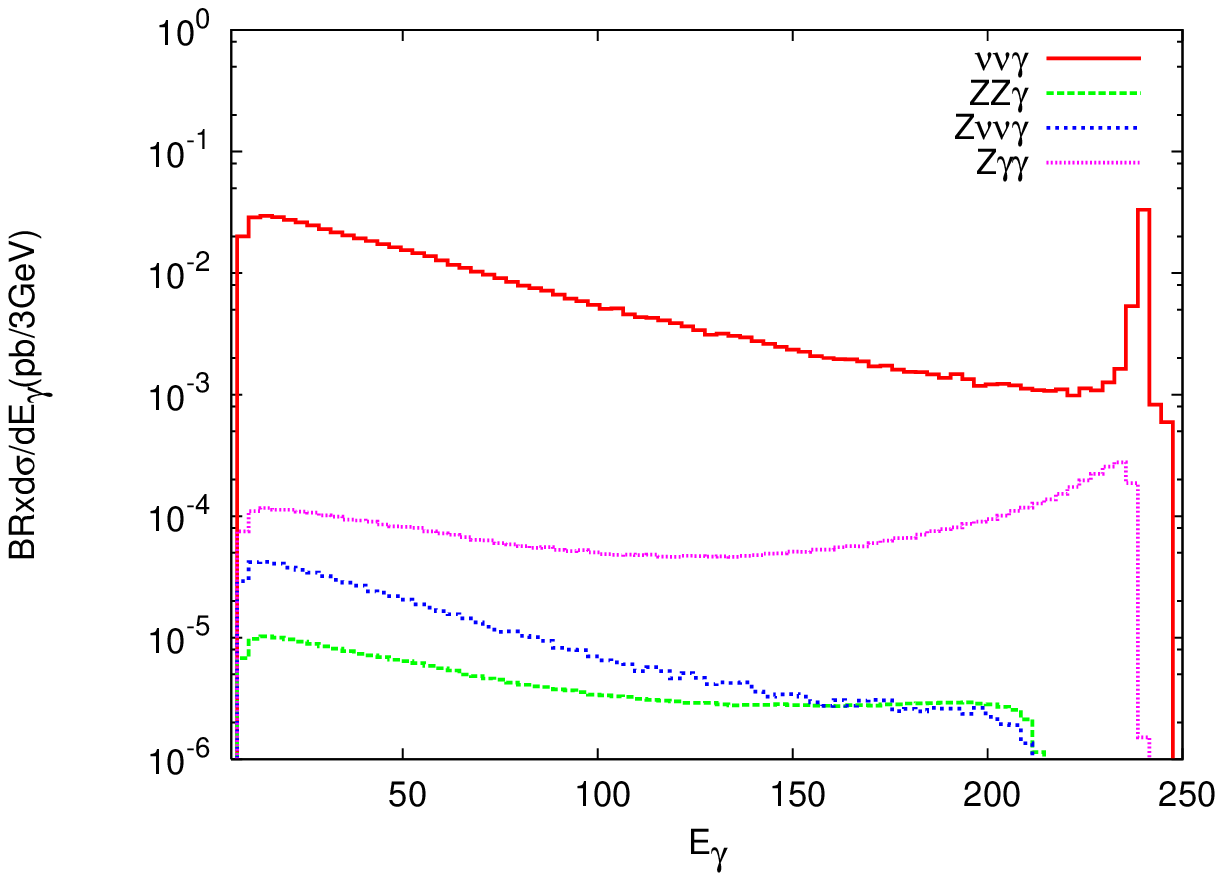}\includegraphics[%
  scale=0.6]{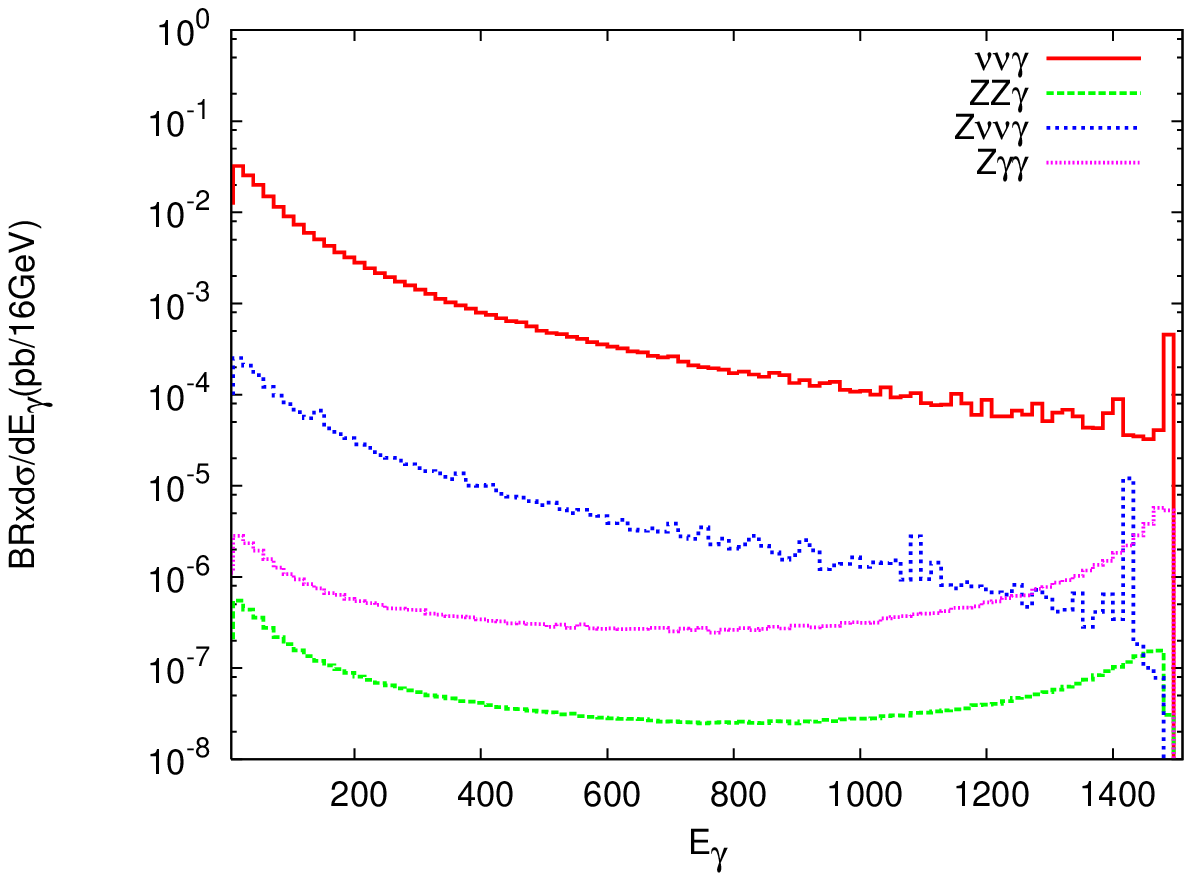}

\caption{Backgrounds contributing to an $"\gamma+nothing"$ analysis. \label{fig7}}
\end{figure}

\begin{table}

\caption{The cross sections (in pb) for $e^{+}e^{-}\to\nu\bar{\nu}\gamma$
with $v/f=0.07$ at $\sqrt{s}=3000$ GeV. The corresponding SM background
gives $\sigma_{B}=3.013$ pb. Here we applied the minimal cuts $E_{\gamma}>10$
GeV, $10^{o}<\theta_{13}<170^{o}$ and $p_{T}>10$ GeV.\label{table4}}

\begin{tabular}{|c|c|c|c|c|c|}
\hline
$\sin\theta$\textbackslash{}$\sin\theta'$&
0.1&
0.3&
0.5&
0.7&
0.9\tabularnewline
\hline
0.1&
3.2502&
3.2093&
3.2206&
3.2359&
3.2311\tabularnewline
\hline
0.3&
4.2023&
3.0384&
3.0505&
3.0578&
3.0614\tabularnewline
\hline
0.5&
10.369&
3.3954&
3.4205&
3.4199&
3.4083\tabularnewline
\hline
0.7&
24.491&
3.1316&
3.1323&
3.1345&
3.1343\tabularnewline
\hline
0.9&
66.303&
3.1130&
3.0709&
3.0768&
3.0722\tabularnewline
\hline
\end{tabular}
\end{table}

\begin{table}

\caption{The cross sections (in fb) for relevant background processes at ILC
and CLIC energies with (w) and without (o) initial state radiation
(ISR) from $e^{+}$ and $e^{-}$ beams. Here, we applied only the
initial cuts.\label{table5}}

\begin{tabular}{|c|c|c|c|c|c|}
\hline
w/o ISR&
$\sigma(\nu\bar{\nu}\gamma)$&
$\sigma(Z\gamma)$&
$ZZ\gamma$&
$Z\nu\bar{\nu}\gamma$&
\tabularnewline
\hline
$\sqrt{s}=0.5$ TeV&
$1843.0/1879.3$&
$2273.0/1730.5$&
$22.94/22.71$&
$10.88/11.76$&
\tabularnewline
\hline
$\sqrt{s}=1$ TeV&
$2372.6/2429.5$&
$582.16/416.13$&
$11.96/11.20$&
$35.73/39.92$&
\tabularnewline
\hline
$\sqrt{s}=3$ TeV&
$2970.4/3012.7$&
$70.03/45.72$&
$3.00/2.63$&
$129.72/133.18$&
\tabularnewline
\hline
$\sqrt{s}=5$ TeV&
$3125.4/3152.2$&
$26.43/16.44$&
$1.44/1.23$&
$174.81/189.04$&
\tabularnewline
\hline
\end{tabular}
\end{table}

We calculate the relevant backgrounds from the reactions $e^{+}e^{-}\to Z\gamma$
($2\to2$) which is the part of $e^{+}e^{-}\to\nu\bar{\nu}\gamma$
($2\to3$) reaction, $e^{+}e^{-}\to ZZ\gamma$ ($2\to3$) and $e^{+}e^{-}\to Z\nu\bar{\nu}\gamma$
($2\to4$) with (w) and without (o) ISR effects at the ILC and CLIC
energies. With the initial cuts we find the background cross sections
as shown in Table \ref{table5}. We see the main contribution to the
background comes from $e^{+}e^{-}\to\nu\bar{\nu}\gamma$ which includes
both $e^{+}e^{-}\to Z\gamma$ ($2\to2$) and $e^{+}e^{-}\to\nu\bar{\nu}\gamma$
($2\to3$ with only $W_{1}$ exchange). Here we take branching ratio
of $Z^{0}\to$\emph{invisible} decay as $20\%$. A background which
cannot be suppressed, comes from the process $e^{+}e^{-}\to\nu\bar{\nu}\nu'\bar{\nu'}\gamma$
with a cross section $\sigma\backsimeq23$ fb. In order to see the
photon energy distribution (between the initial cuts and kinematical
cuts) of these backgrounds in the $"\gamma+nothing"$ analysis we
show differential cross sections multiplied by corresponding branching
ratios in Fig. \ref{fig7} at the center of mass energies $\sqrt{s}=0.5$
TeV and $\sqrt{s}=3$ TeV. Here, we assume lepton universality, and
calculate the cross sections to give an idea about the magnitude of
the background considered. In general, applying some strict cuts around
the resonance regions and by making an optimization for $S/B$ ratio,
the measurements can also be improved, provided that the LHC measures
the masses of the extra gauge bosons predicted by the LHM.

For a given center of mass energy we can determine the contributions
from new gauge bosons in different parameter regions: one is the resonant
region where a peak in the distribution is obtained for some certain
values of the parameters $s,s'$ and $f$; second is non-resonant
region where the parameter scans can be performed over a wide range;
third is the decoupling region ($c'=\sqrt{2/5}$) where the coupling
of $Z_{3}$ to fermions vanishes, here there is also another approach
that the mass of the new gauge boson can be taken infinitely heavy.
We show the results for the mentioned cases during our analysis.

\begin{figure}
\begin{center}\includegraphics[%
  scale=0.6]{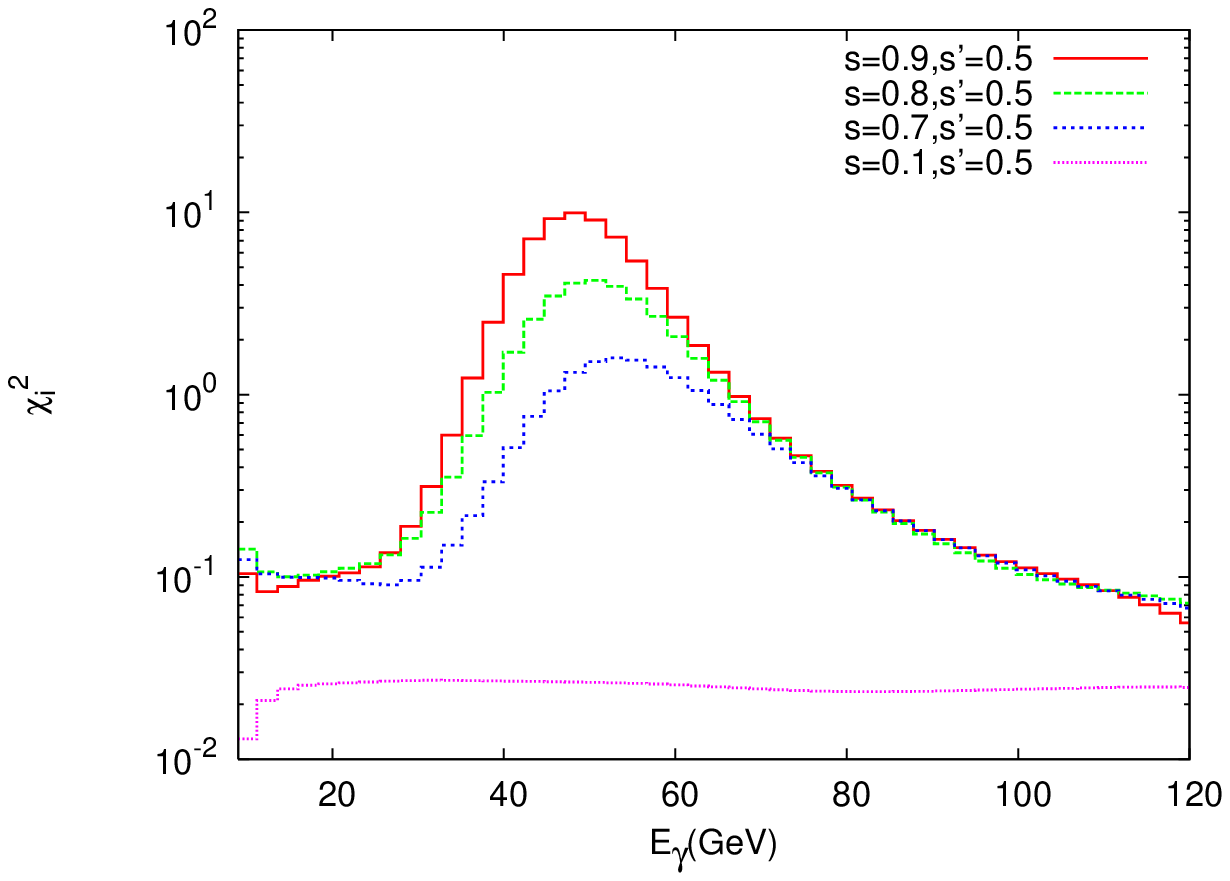}\includegraphics[%
  scale=0.6]{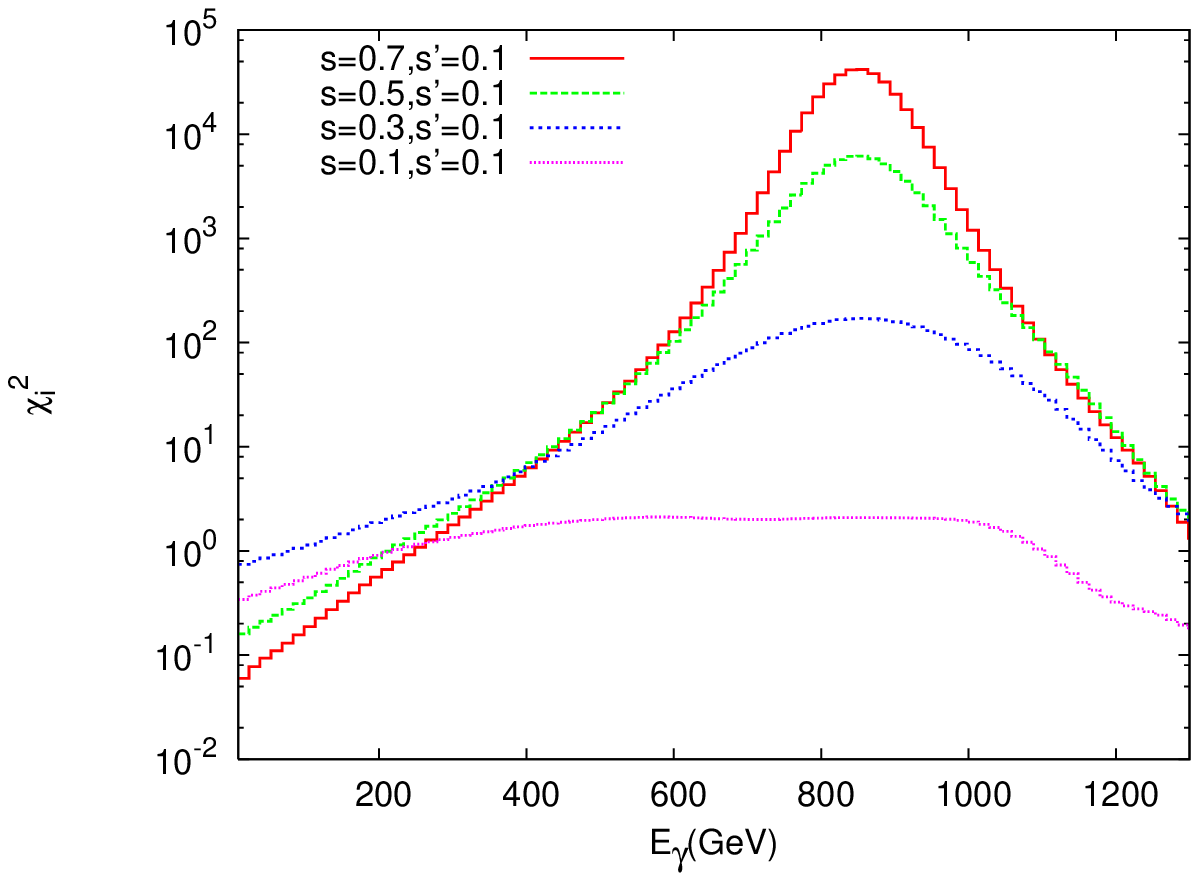}\end{center}

\caption{The $\chi_{i}^{2}$ distribution depending on the energy bins $i$
for different LHM mixing parameters at ILC with $\sqrt{s}=500$GeV
(left) and CLIC with $\sqrt{s}=3$ TeV (right) , here we assume $L_{int}=100$
fb$^{-1}$ and $v/f=0.1$. \label{fig8}}
\end{figure}

In order to obtain the discovery limits of the LHM parameters we perform
the $\chi^{2}$ analysis. We calculate the $\chi^{2}$ distribution
as

\begin{equation}
\chi^{2}={\displaystyle \sum_{k=1}^{n}\left(\frac{\frac{d\sigma^{k}}{dE_{\gamma}}(\mbox{LHM})-\frac{d\sigma^{k}}{dE_{\gamma}}(\mbox{SM})}{\delta\frac{d\sigma^{k}}{dE_{\gamma}}(\mbox{SM})}\right)^{2}}\label{eq:17}\end{equation}
where $\delta d\sigma^{k}/dE_{\gamma}$ is the error on the measurement
including statistical and systematical errors added in quadrature.
As we already noted that the backgrounds are much smaller than the
signal, we expect the statistical errors in the SM backgrounds would
be smaller than the systematic errors including detector and $e^{-}/e^{+}$
beam uncertainties. Here, we considered a systematic error $\delta_{sys}=5\%$
for a measurement. This may be an overestimate, however, if improved
the constraints can be relaxed and benefit from the advantage of high
luminosity. The differential cross section depends on the model parameters
$s,s'$ and $f$. We may assume that the LHC would have determined
the mass of the extra gauge bosons relatively well, to the order of
a few percent. Thus we can fix $m_{Z_{i}}$and perform a two-parameter
scan. We calculate $\chi^{2}$ at every point of $s,s'$. In this
case $\chi^{2}=\chi_{\min}^{2}+C$. The constraint on the parameters
with $95\%$ C.L. can be obtained at the ILC and CLIC energies by
requiring $C=5.99$ for two free parameters. In calculating the $\chi^{2}$
for $d\sigma/dE_{\gamma}$we have used equal sized bins in the range
$E_{\gamma}^{min}<E_{\gamma}<E_{\gamma}^{max}$ where the upper limit
is taken as the kinematical limit for the photon energy. The most
sensitive results can be obtained for $s'=0.5(0.1)$ at the center
of mass energy $\sqrt{s}=0.5(3)$ TeV as shown in Fig. \ref{fig8}.
The $\chi_{i}^{2}$ distributions versus the photon energy bins show
peaks shifted to the right depending on lower $s$ and lower $f$
values. Here we have used $v/f=0.1$ and $0.07$ for the ILC and CLIC
energies, respectively.

\begin{figure}
\includegraphics[%
  scale=0.7]{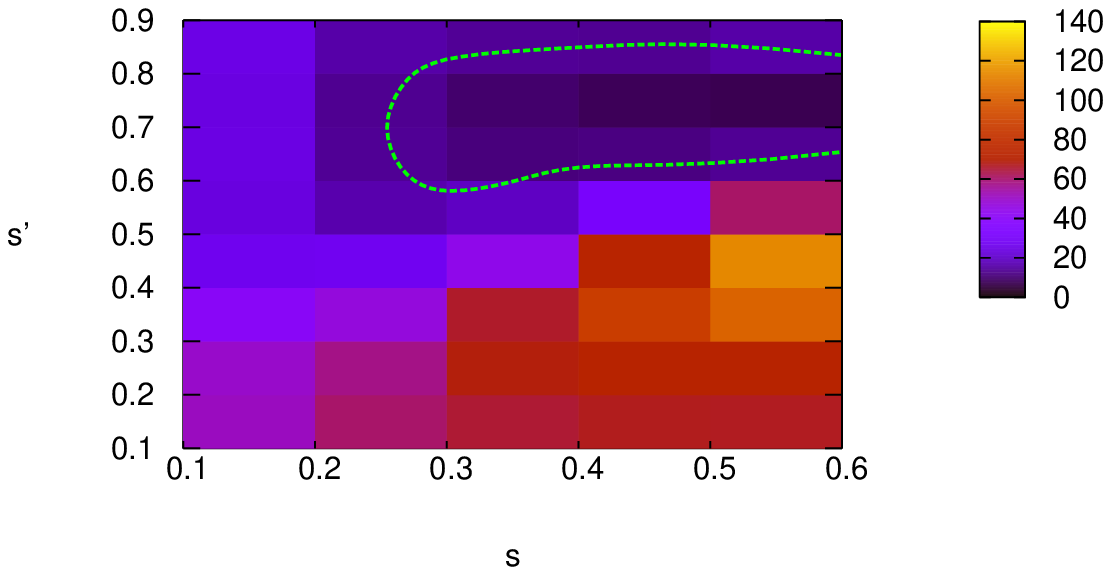} \includegraphics[%
  scale=0.7]{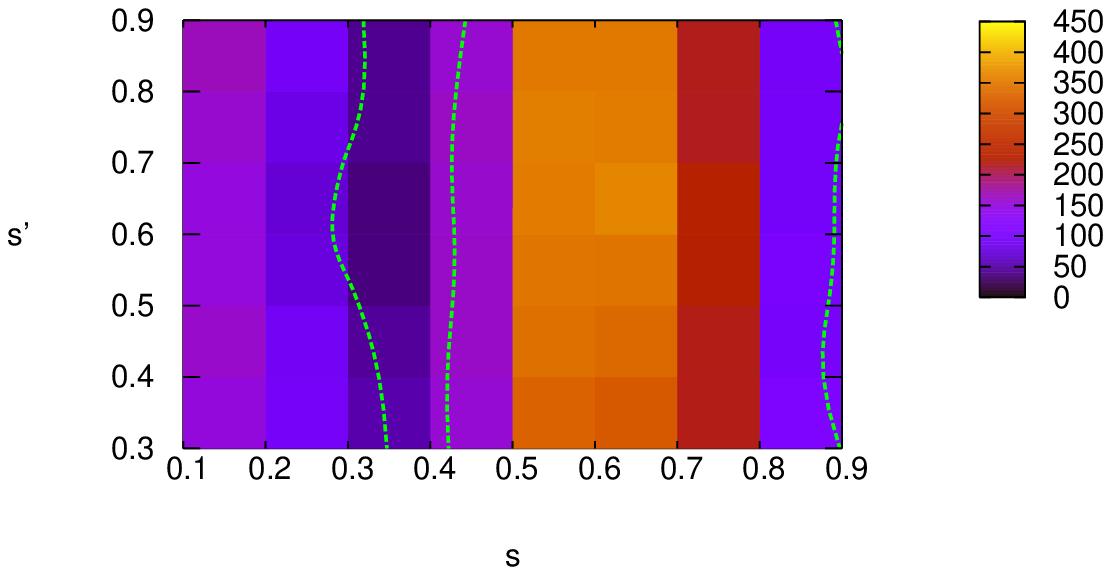}

\caption{The density plot and the contour lines with $95\%$ C.L. for the
search reach in the parameter space ($s,s'$) with $v/f=0.1$ (left)
and $v/f=0.07$ (right) at ILC (left) and CLIC (right) energies. \label{fig9}}
\end{figure}

In Fig. \ref{fig9} we present the constraints on mixing parameters
$s,s'$ in a density plot. For the $Z_{3}$ search at the ILC energies
with $L_{int}=100$ fb$^{-1}$ most of the $s,s'$ parameter space
can be discoverd. A contour line for the constrained parameter space
($s,s'$) is also shown on the plot. We may exclude the region with
$0.6<s'<0.8$, $0.25<s<0.9$ by this analysis at $\sqrt{s}=0.5$ TeV.
When the systematic error is not included, the shape of the plot is
luminosity dependent, even for a low luminosity as $L_{int}\sim10^{3}$
pb$^{-1}$ only the decoupling region ($s'=\sqrt{3/5}$) remains dark
(not accessible) in this plot. At higher center of mass energies different
parameter regions can be constrained. The resonance regions deserve
special attention at the ILC and CLIC energies. Because the highest
sensitivity to new physics is obtained in this region. Taking $s'=0.5$
we can probe the $Z_{3}$ signal for the interested range of $s=0.5-0.9$
and $f=0.5-2.7$ TeV at $\sqrt{s}=0.5$ TeV and $L_{int}=100$ fb$^{-1}$
. For the CLIC at $\sqrt{s}=3$ TeV and $L_{int}=100$ fb$^{-1}$,
and taking the mixing parameter $s'=0.1$, we can probe the resonance
peaks between the scale $f=1-3.7$ TeV for almost all range of $s$.
The extra gauge boson signals of LHM can be measured for almost all
interested range of $s,s'$ except $0.3<s<0.4$ at CLIC with a projected
luminosity $L_{int}=100$ fb$^{-1}.$

\begin{figure}
\includegraphics[%
  scale=0.6]{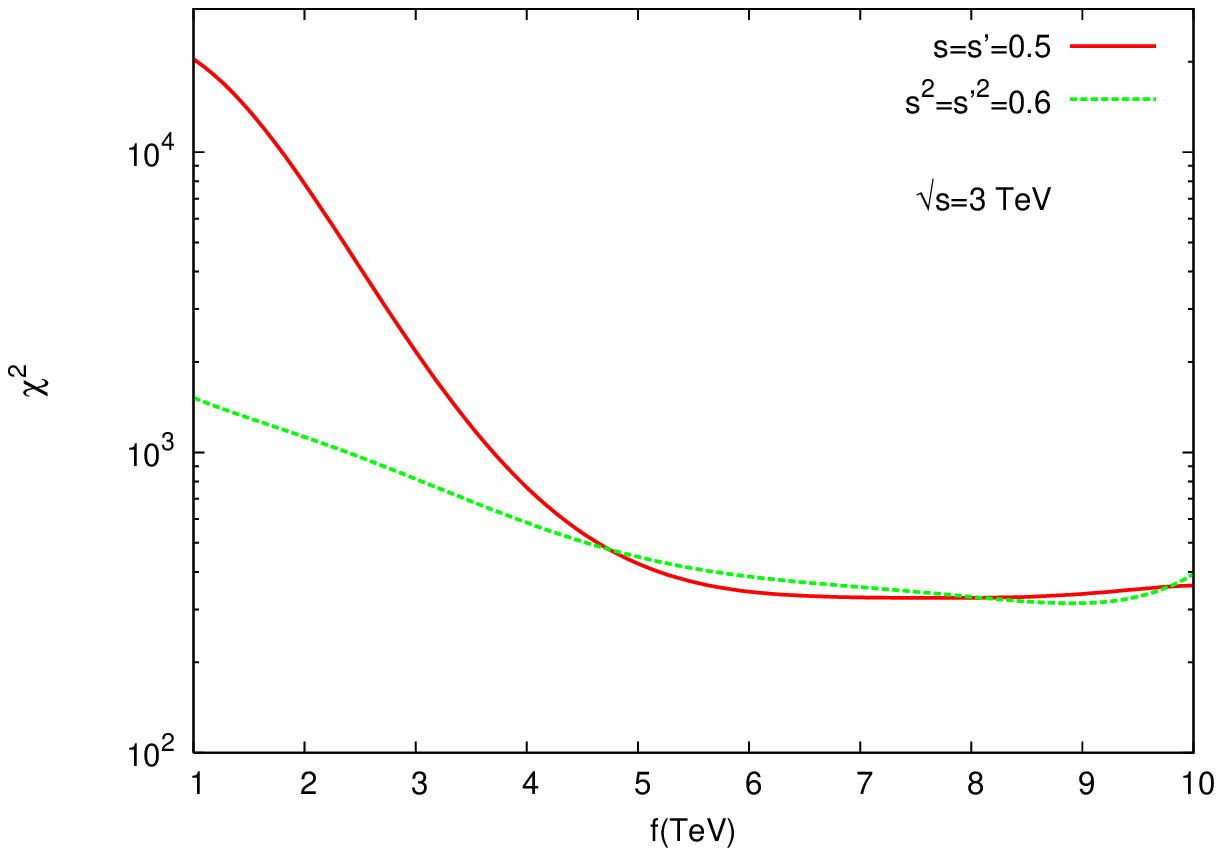} \includegraphics[%
  scale=0.6]{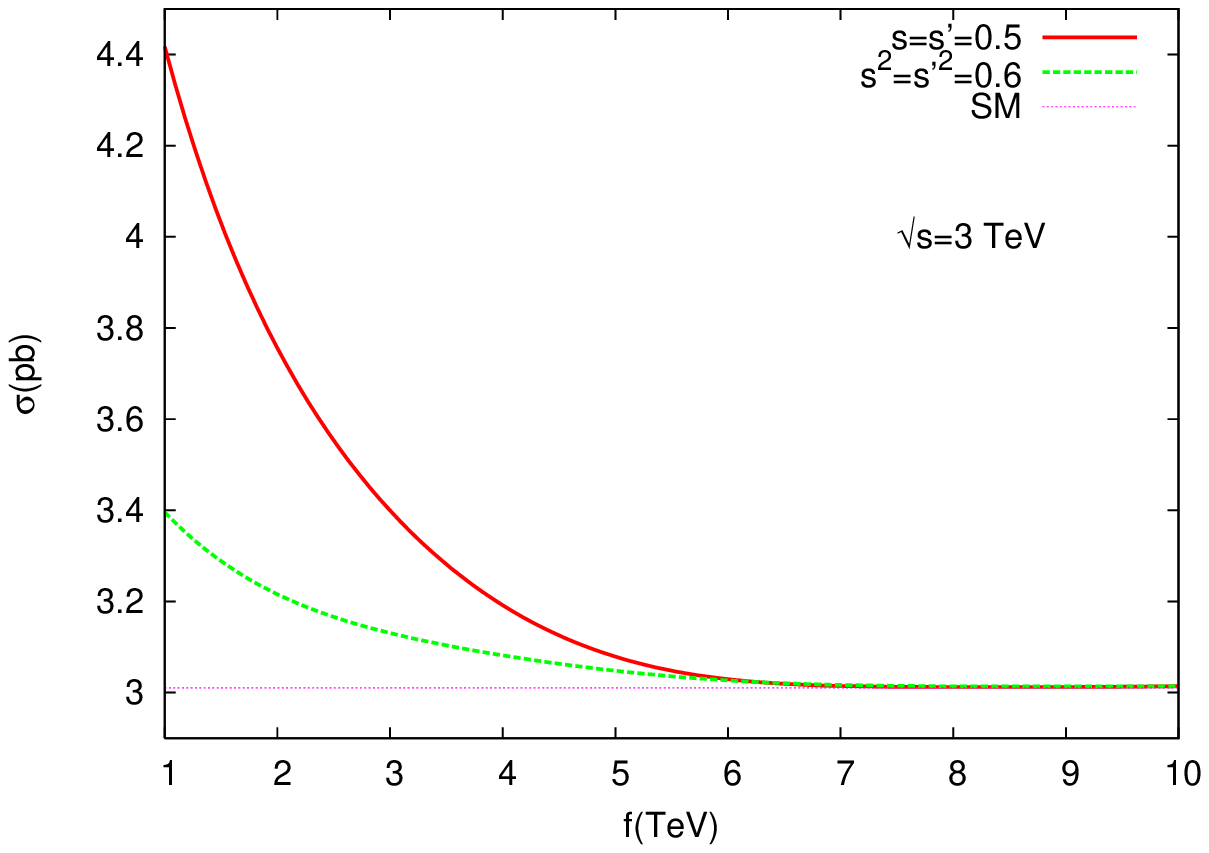}

\caption{The $\chi^{2}$ plot (left), and total cross sections for the LHM
signal and SM background (right) versus the scale $f$ for fixed values
of the parameters $s,s'=0.5$ and $s,s'=\sqrt{3/5}$ at CLIC with
$\sqrt{s}=3$ TeV, $L_{int}=100$ fb$^{-1}$. \label{fig10}}
\end{figure}

We continue our analysis with higher values of the scale $f$ as $f\gtrsim4$
TeV, and we would like to determine the accuracy of the parameter
measurements using a $\chi^{2}$ analysis. The best discovery limit
is obtained using the observable $d\sigma/dE_{\gamma}$. We calculate
the $\chi_{\min}^{2}$, and determine the discovery region corresponding
to $\chi^{2}<\chi_{\min}^{2}+2.69$ for one free parameter. As the
references we take the parameters $s,s'=1/2$ and $s,s'=\sqrt{3/5}$,
first is arbitrary but the latter corresponds to decoupling the $Z_{3}$
from the leptonic current. If the masses of extra gauge bosons can
not be measured at the LHC, we may need to scan parameter $f$ at
higher energies. In Fig. \ref{fig10}, for the CLIC energies we depict
the $\chi^{2}$ plot versus the scale $f$ for fixed values of $s$
and $s'$. We also show the signal and background cross sections versus
$f$. Based on the analysis mentioned above, the parameter $f$ can
be reached up to 6 TeV at CLIC with $\sqrt{s}=3$ TeV. We can measure
the scale $f$ (or the mass of heavy gauge boson) with an error of
$5\%$. This limit enhances when we take into account smaller systematic
errors for a measurement.

\section{Conclusions}

In this work, we have studied the sensitivity of the process $e^{+}e^{-}\to\nu\bar{\nu}\gamma$
to the extra gauge bosons $Z_{2},Z_{3}$ and $W_{2}^{\pm}$ in the
framework of the little Higgs model. The search reach of the ILC (operating
at $\sqrt{s}=0.5$ TeV and $L_{int}=100$ fb$^{-1}$ for one year)
and CLIC (when operating at $\sqrt{s}=3$ TeV, and $L_{int}=100$
fb$^{-1}$) covers a wide range of parameter space where this model
relevant to the hierarchy. For the parameter space where the resonances
occur ($s'=0.5(0.1)$) by scanning the parameter $s$, we can access
the range for scale $f=0.5-2.7$ ($1-3.7$) TeV at $\sqrt{s}=0.5$
($3$) TeV, respectively. If the scale $f$ is larger than $f\gtrsim4$
TeV, a sensitivity to the parameters of LHM could be reached with
a detailed MC including detector and beam luminosity/energy uncertainty
effects.

Finally, the ILC and CLIC with high luminosity have a high search
potential for different regions of parameter space of the LHM. Analysis
of $e^{+}e^{-}\to\nu\bar{\nu}\gamma$ process can give valuable information
about the LHM and it can serve a clean environment for precise determination
its parameters. The measurements with small systematic errors are
needed to have desired sensitivity for the new physics parameters.
Even for the cases in which search reach for extra gauge bosons in
this process is not competitive with the potential of the LHC, the
measurements at linear colliders can also provide detailed information
on extra gauge bosons which complements the results from the LHC.

\begin{acknowledgments}
The work of O.C. was supported in part by the State Planning Organization
(DPT) under the grants no DPT-2006K-120470 and in part by the Turkish
Atomic Energy Authority (TAEA) under the grants no VII-B.04.DPT.1.05.
\end{acknowledgments}

\end{document}